\documentclass[10pt,english]{IEEEtran}
\usepackage{graphicx,myPack,myMathPack,nicefrac,mathtools,algorithm2e}
\usepackage[caption=false,font=footnotesize]{subfig}

\sclidx{\timeIdx}{t}
\sclidx{\nodeIdx}{n}
\gset{\nodeSet}{N}
\sclidx{\numNodes}{N}

\sclidx{\numRounds}{R}
\sclidx{\roundIdx}{r}
\sclidx{\iterIdx}{k}
\scalar{\iterIdxb}{\ell}

\sclidx{\nodeIdxFunc}{w}

\rvect{\globState}{S}
\rscalar{\locState}{S}
\vect{\globStatei}{i}
\vect{\globStatej}{j}
\newcommand{\globStateSet}{\boldsymbol{\mathcal{S}}}
\gset{\locStateSet}{S}
\scalar{\locStatei}{i}
\scalar{\locStatej}{j}
\scalar{\locStatek}{k}

\rscalar{\action}{A}
\scalar{\actiona}{a}
\gset{\actionSet}{A}

\func{\rewardFunc}{R}
\func{\valueFunc}{V}
\func{\controlMap}{c}
\rscalar{\jmessage}{M}
\scalar{\possibleMess}{m}
\scalar{\messOrderMap}{\mu}
\func{\locStateEncoder}{q}
\vecfunc{\locStateEncoderVec}{q}

\gset{\binStrings}{B}
\gset{\feasSet}{F}

\scalar{\controlRateLagrMul}{\lambda}
\scalar{\discountFactor}{\gamma}

\scalar{\mdpTransDist}{p}

\scalar{\entropy}{H}

\gset{\charGraph}{G}
\gset{\charGraphNodes}{V}
\gset{\charGraphEdges}{E}

\gset{\allColorings}{R}
\func{\jcoloring}{r}

\gset{\candControlMapSet}{C}

\scalar{\minRateNoInt}{R_{\downarrow,ni}}
\scalar{\rateReduc}{\rho}
\newcommand{\probDistrSet}{\mathcal{P}}

\newcommand{\probDistrVec}{\boldsymbol{p}}
\newcommand{\probDistr}{\boldsymbol{p}}

\newcommand{\Reals}{\mathbb{R}}

\newcommand{\upperConcave}{\textrm{concavify}}

\newcommand{\minRateWI}{R^{\downarrow,wi}}

\newcommand{\stateOrderer}{\ell}
\newcommand{\transMatrix}{\mathbf{P}}

\newcommand{\initStateDistrV}{\boldsymbol{\pi}}

\newcommand{\identityMatrix}{\mathbb{I}}

\title{A Framework for Rate Efficient Control of Distributed Discrete Systems}
\author{Jie Ren, Solmaz Torabi, John MacLaren Walsh}
\date{June 1, 2016}

\begin{document}
\maketitle

\begin{abstract}
A key issue in the control of distributed discrete systems modeled as Markov decisions processes, is that often the state of the system is not directly observable at any single location in the system.  The participants in the control scheme must share information with one another regarding the state of the system in order to collectively make informed control decisions, but this information sharing can be costly.  Harnessing recent results from information theory regarding distributed function computation, in this paper we derive, for several information sharing model structures, the minimum amount of control information that must be exchanged to enable local participants to derive the same control decisions as an imaginary omniscient controller having full knowledge of the global state.  Incorporating consideration for this amount of information that must be exchanged into the reward enables one to trade the competing objectives of minimizing this control information exchange and maximizing the performance of the controller.  An alternating optimization framework is then provided to help find the efficient controllers and messaging schemes.  A series of running examples from wireless resource allocation illustrate the ideas and design tradeoffs.
\end{abstract}

\section{Introduction}
The framework of Markov decision processes (MDP) \cite{Bellman1957MDP,MDP} provides a principled method for the optimal design of controllers for discrete systems.  By solving a Bellman's equation, for example through either a value or policy iteration, one derives a control mapping assigning to each possible state of the system a control action to take, in a manner that maximizes a long run discounted expected reward.   Of increasing interest, however, are those discrete systems which are decentralized or distributed, in the sense that no single participant in the system has full access to the global system state, but rather this global state is the concatenation of a series of local states, each of which are directly observed at different locations.  

For instance, a series of agents may each observe their own local state, and have a set of local actions to choose from \cite{Hsu1982,Ho1972,Sandell1978}, and the desire may be to design individual local controllers.  The fact that the global system state is not available at any location either requires sufficient information to be exchanged, either through ordinary communications or through the system's state \cite{park2011network,park2013network}, to remedy the situation, or a modification of the control framework.  One way to address in part this distributed knowledge of network state is to use the framework of partially observable Markov decision processes (POMDP) to synthesize controllers.  In general optimal control of a POMDP requires the controller to maintain probabilistic belief-states about the current system state based on all previous control actions and all previous observations, then assigning a control action based on these belief states.  Thus a key issue in the selection/design of POMDPs are problem structures which enable simple forms for this control action \cite{VikramPOMDP}.  As they are even more complex, solving general decentralized multiagent (PO)MDPs are NEXP-complete \cite{Bernstein2000}, and a rich literature, e.g. \cite{Boutilier1996MMDP,Nair2003DecPOMDP,Chades2002,Szer2005,Olienhoek2008,Seuken2008} and many others, have addressed finding approximate solutions.  

A key issue in the literature about decentralized and/or distributed MDPs is the role and amount of communication and coordination.  Relationships between communication and control have been established in several contexts in the literature.  Information theoretic limits have been incorporated into the classical case of a single observer remotely controlling a linear system through a noisy channel \cite{Tatikonda1,Tatikonda2},  with a focus on determining the minimum rate required to achieve control objectives \cite{Tatikonda2} \cite{Borkar1997}, \cite{Wong1997}.   In this vein, \cite{sahai2006necessity} proposes the notion of anytime capacity and gives a necessary and sufficient condition on communication reliability needed over a noisy channel to stabilize an unstable scalar linear process when the observer has access to noiseless feedback of the channel output. Building on these ideas, \cite{sukhavasi2011anytime,sukhavasi2011linear,sukhavasi2016linear} provide explicit construction of anytime reliable codes with efficient encoding and decoding over noisy channels.  Shifting to a decentralized control context, deep connections between communications between controllers through the system state and network coding have recently been investigated \cite{park2011algebraic,park2011network,park2013network}.

Despite this long standing interest between relationships between communication and control, the role of information theoretic limits when state observations from a MDP are distributed has been less well developed in the literature.  Here, the literature studying communication and coordination has not made use of related ideas from multiterminal information theory to compute communication cost.  \cite{Gmytrasiewicz2000}, for instance, considers a multi-agent coordination with agent decisions made in a self-interested environment, while \cite{Pynadath2002COM-MTDP} discusses the computational complexity of finding optimal decisions in a communicative multi-agent team decision problem (COM-MTDP) along the dimension of communication cost.  Similarly \cite{xuan2000communication, Xuan1, Xuan2}, recognize that communication incurs a cost in the global reward function, and show that whether or not to communicate is also a decision to make.  However, none of these models in \cite{Gmytrasiewicz2000,Pynadath2002COM-MTDP,xuan2000communication, Xuan1, Xuan2} make use of relevant information theoretic limits when computing a communication cost.  Part of the reason that information theoretic limits have not been fully brought to bear in the distributed MDP is that the limits for the relevant models, for instance for distributed \cite{FunctionCompression,FunctionComputing} and/or interactive function computation \cite{Pra:Rat,InteractiveCollocated}, have been only somewhat recently derived.  Bearing this in mind, this paper aims to harness information theoretic limits and coding designs that approach them, to help synthesis efficient control schemes for a distributed MDP.  

In particular, in Section \ref{sec:mdp} we consider a MDP model in which the state can be considered a vector of local states, each of which are directly observed at a different location.  If the goal is to have this distributed MDP operate in the same manner as an omniscient MDP having full access to this global state, then a natural way to achieve this is through the exchange of state information between the participants.  This information exchange, however, can be costly, and furthermore, in order to achieve the same performance, i.e. simulate, an omniscient controller, all that is truly necessary is that the system make the same decisions.  As such, it can be desirable to determine the minimal amount of information that must be exchanged in order to enable the distributed system to learn the decisions the centralized omniscient controller would take.  Lower bounds on this minimal amount of control information necessary are provided in Section \ref{sec:min-communication-cost} using recent results from information theory regarding quantization, both interactive and non-interactive, for decentralized function computation.  Alternatively as described in Section \ref{sec:iterative}, one may wish to incorporate the cost of the communication into the reward, enabling the control designer to trade the cost of the communication for the performance of the controller.  In this case, the messaging scheme and controller must be designed simultaneously, and in order to provide candidate solutions to the associated optimization, we present in Section \ref{sec:alg} an alternating optimization based method guaranteed to yield a sequence of rewards which converges.  When the the messaging scheme and control map selected by this scheme converges, we prove that it must lie at a Nash equilibrium of the total reward, which combines the reward of the controller with the cost of the communication.  Several examples drawn from the design of downlink resource controllers for wireless networks throughout the manuscript demonstrate the ideas.

\section{Omniscient Control of a Distributed Markov Decision Process}
\label{sec:mdp}
Consider a distributed discrete stochastic control system modeled a Markov Decision Process (MDP) for whom the global system state at time $\timeIdx$, $\globState_{\timeIdx} \in \globStateSet$, is itself a vector composed of a series of local states $\locState_{\nodeIdx,\timeIdx}\in\locStateSet_{\nodeIdx}$ each observed at one network node $\nodeIdx\in\nodeSet = \{ 1,\ldots, \numNodes\}$, so that
\begin{equation} \label{eq:distributed-states}
\globState_{\timeIdx} = \left[ \locState_{\nodeIdx,\timeIdx} | \nodeIdx \in \nodeSet \right] \quad \textrm{and} \quad \globStateSet = \locStateSet_{1}\times \cdots \times \locStateSet_{\numNodes}
\end{equation}
The global network state $\globState_{\timeIdx}$ evolves according to a Markov Chain whose transition matrix at time $\timeIdx$ is selected by the control action $\action_{\timeIdx}$
\begin{equation}
\mathbb{P}\left[ \globState_{\timeIdx+1}=\globStatej \left| \globState_{\timeIdx}=\globStatei,\action_{\timeIdx} = \actiona \right. \right] = \mdpTransDist_{\actiona }(\globStatei,\globStatej) \quad \forall \globStatei,\globStatej\in\globStateSet.
\end{equation}
Additionally, there is a reward function $\rewardFunc_{\actiona}(\globStatei,\globStatej)$ indicating the payment obtained when the global system state transitions from $\globStatei$ to $\globStatej$ after action $\actiona$ is taken.

An omniscient controller having access to the series of global states $\globState_{\timeIdx}, \timeIdx \in \mathbb{N}$ would select the actions $\action_{\timeIdx}$ maximizing the total discounted expected reward
\begin{equation}\label{eq:total-discounted-reward}
\min_{\controlMap: \globStateSet\rightarrow \actionSet}  \sum_{\timeIdx=0}^{\infty} \discountFactor^{\timeIdx} \mathbb{E}[\rewardFunc_{\action_{\timeIdx}}(\globState_{\timeIdx},\globState_{\timeIdx+1})] .
\end{equation}
The argument to the solution to this optimization is a mapping $\controlMap:\globStateSet \rightarrow \actionSet$ assigning to each state the optimum action to take, so that the optimal $\action_{\timeIdx}^* = \controlMap(\globState_{\timeIdx})$.  Bellman's equation states that the solution to this optimization must solve the following system of equations
\begin{equation}\label{eq:optValue}
\valueFunc^*(\globStatei) = \sum_{\globStatej \in\globStateSet} \mdpTransDist_{\controlMap^*(\globStatei)}(\globStatei,\globStatej) \left[ \rewardFunc_{\controlMap^*(\globStatei)}(\globStatei,\globStatej) + \discountFactor \valueFunc^* (\globStatej)\right] \quad \forall \globStatei \in \globStateSet 
\end{equation}
\begin{equation}\label{eq:optControl}
\begin{aligned}
\controlMap^*&(\globStatei) = \\
&\underset{\actiona\in\actionSet}{\arg\max}\sum_{\globStatej \in\globStateSet} \mdpTransDist_{\actiona}(\globStatei,\globStatej) \left[ \rewardFunc_{\actiona}(\globStatei,\globStatej) + \discountFactor \valueFunc^* (\globStatej)\right]\quad \forall \globStatei \in \globStateSet\\ 
\end{aligned}
\end{equation}

The solution to this simultaneous system of equations, and the associated optimal control mapping, can be found by first determining the limit $\valueFunc^* = \lim_{\iterIdx \rightarrow \infty} \valueFunc_{\iterIdx}$ of the following value iteration
\begin{equation}
\valueFunc_{\iterIdx+1}(\globStatei) = \max_{\actiona\in\actionSet} \sum_{\globStatej \in\globStateSet} \mdpTransDist_{\actiona}(\globStatei,\globStatej) \left[ \rewardFunc_{\actiona}(\globStatei,\globStatej) + \discountFactor \valueFunc_{\iterIdx} (\globStatej)\right] \quad \forall \globStatei \in \globStateSet
\end{equation}
then solving for the control policy via
\begin{equation}
\begin{aligned}
\controlMap^{\ast}&(\globStatei) =\\
& \underset{\actiona \in\actionSet}{\arg \max}\sum_{\globStatej \in\globStateSet} \mdpTransDist_{\actiona}(\globStatei,\globStatej) \left[ \rewardFunc_{\actiona}(\globStatei,\globStatej) + \discountFactor \valueFunc^* (\globStatej)\right] \quad \forall \globStatei\in\globStateSet.\\
\end{aligned}
\end{equation}

Alternatively, one can utilize a policy iteration, which performs a recursion in which first 
\begin{equation}
\controlMap_{\iterIdx}(\globStatei) = \underset{\actiona \in\actionSet}{\arg\max} \sum_{\globStatej \in\globStateSet} \mdpTransDist_{\actiona}(\globStatei,\globStatej) \left[ \rewardFunc_{\actiona}(\globStatei,\globStatej) + \discountFactor  \valueFunc_{\iterIdx} (\globStatej)\right] \quad \forall \globStatei\in\globStateSet
\end{equation}
is solved, followed by a solution of the linear system
\begin{equation}
\begin{aligned}
\valueFunc_{\iterIdx+1}&(\globStatei) =  \\
&\sum_{\globStatej \in\globStateSet} \mdpTransDist_{\controlMap_{\iterIdx}(\globStatei)}(\globStatei,\globStatej) \left[ \rewardFunc_{\controlMap_{\iterIdx}(\globStatei)}(\globStatei,\globStatej) + \discountFactor \valueFunc_{\iterIdx+1} (\globStatej)\right] \quad \forall \globStatei \in \globStateSet\\
\end{aligned}
\end{equation}
for $\valueFunc_{\iterIdx+1}(\cdot)$.
until the control mapping can be selected to remain the same under the update, at which point the iteration ceases, see, e.g. \cite{DynamicProgramming}\cite{VikramPOMDP}.

Note that in many problems, for a given $\globStatei$, there is more than one choice for $\controlMap^*(\globStatei)$ achieving the maximum in (\ref{eq:optControl}).  In this instance, one can derive a set of candidate (omniscient) control functions $\candControlMapSet$ which are those $\controlMap^*: \globStateSet \rightarrow \actionSet$ obeying the constraints
\begin{equation}\label{eq:cand-optControl}
\begin{aligned}
\controlMap^*&(\globStatei) \in\\
& \underset{\actiona\in\actionSet}{\arg\max} \sum_{\globStatej \in\globStateSet} \mdpTransDist_{\actiona}(\globStatei,\globStatej) \left[ \rewardFunc_{\controlMap^*(\globStatei)}(\globStatei,\globStatej) + \discountFactor \valueFunc^* (\globStatej)\right],\ \forall \globStatei \in \globStateSet \\
\end{aligned}
\end{equation}
each of which achieves the maximum long run expected reward.

\begin{example}[Downlink Wireless Resource Allocation]
\label{exmp:resourceallocation1}
An example of a practical problem having the structure outlined by (\ref{eq:optValue}) and (\ref{eq:optControl}) is that of distributing resources on a wireless downlink.  Time is slotted.  There is a basestation which has a shared buffer containing individual information that must be sent to a series of users, with the amount waiting for a user $\nodeIdx \in \{1,\ldots,\numNodes\}$ during time slot $\timeIdx \in \mathbb{N}$ being denoted by $B_{\nodeIdx,\timeIdx}$.  Collectively, these buffer sizes form the local state at the basestation $\locState_{\numNodes+1,\timeIdx} = (B_{1,\timeIdx},\ldots,B_{\numNodes,\timeIdx})$.  Each user has a channel state $\locState_{\nodeIdx,\timeIdx}$, indicating how much information can be reliably transmitted to this user during the present timeslot, which evolves independently of other users channel states according to a Markov chain with transition distribution $\mdpTransDist(\locStatei_{\nodeIdx},\locStatej_{\nodeIdx}) = \mathbb{P}[\locState_{\nodeIdx,\timeIdx+1}=\locStatej_{\nodeIdx}| \locState_{\nodeIdx,\timeIdx}=\locStatei_{\nodeIdx}]$.  During each time slot $\timeIdx$, a random amount of additional traffic $X_{\nodeIdx,\timeIdx}$ arrives destined for each user $\nodeIdx \in \{1,\ldots, \numNodes\}$ at the basestation's buffer, independently of other time slots and previous arrivals.  If the basestation's buffer can not accommodate this traffic, it is dropped.  

During each time slot, it must be decided which of the users to give the resource to, and thus this forms the action in the MDP, $\actionSet = \{1,\ldots,\numNodes\}$.  Additionally, both the users and the basestation must know the outcome of this decision.  After the user to transmit to is selected, an amount of their traffic that is the minimum between their capacity during the slot and the amount of traffic waiting for them in the buffer will be successfully transmitted to them and removed from the buffer, yielding the Markov chain dynamics
\begin{equation}
\locState_{\numNodes+1,\timeIdx+1} = D(\locState_{\numNodes+1,\timeIdx} - \boldsymbol{T}(\action_{\timeIdx},\globState_{\timeIdx}) + \boldsymbol{X}_{\timeIdx+1},\boldsymbol{X}_{\timeIdx+1}),
\end{equation}
where $\boldsymbol{X}_{\timeIdx} = [X_{1,\timeIdx},\ldots,X_{\numNodes,\timeIdx}]$ is the amount of new traffic arriving during the time slot $\timeIdx$ for each user, and the vector $\boldsymbol{T}(\action_{\timeIdx},\globState_{\timeIdx-1})$ has $\nodeIdx$th element
\begin{equation}
T_{\nodeIdx}(\action_{\timeIdx},\globState_{\timeIdx-1}) = \left\{ \begin{array}{cc} \min\{ \locState_{\action_{\timeIdx},\timeIdx}, B_{\action_{\timeIdx},\timeIdx} \} & \nodeIdx = \action_{\timeIdx} \\ 0 & \textrm{otherwise} \end{array} \right.
\end{equation}
while the function $D$ performs the package dropping process when the arriving traffic can not be accommodated in the buffer.  Note that this assumption, that for this amount of information to be successfully transmitted and received, all the users and basestation need to know is who to schedule, is consistent with assumption that the physical layer below the scheduler uses a rateless code with feedback, which can be closely approximated with hybrid ARQ \cite{LubyLT,VariableRateCC,RatelessAWGN}.

When making the decision of who to schedule, several important metrics can be considered, and thereby combined, into the reward.  A very natural metric is the throughput, which measures how much information is transmitted, summed over all the users.  This gives the reward function
\begin{equation} \label{eq:throughput}
\rewardFunc_{\actiona}(\globState_{\timeIdx},\globState_{\timeIdx+1}) = \min(\locState_{\actiona,\timeIdx},B_{\actiona,\timeIdx})
\end{equation}
Other metrics such as the average or maximum delay a users traffic experiences are also reasonable metrics which can be incorporated into the reward function, for instance by adding them together with rates that can trade them for one another.

With the selected metrics included in the reward function, the MDP framework gives a formal way of deciding who to allocate the resources to during each time slot.  A series of examples throughout the rest of the paper will find this optimal controller and investigate properties, such as how much information must be exchanged in order to perform it.
\end{example}

\subsection{Simulating the Omniscient MDP via Information Exchange}
However, as the system is distributed, no single node is given access to the global network state, and control must be performed by the nodes learning which action to take through some sort of communication enabled strategy.  Additionally, we will introduce the constraint that an observer not given access to any of the local states, but rather accessing only all of the control messages the nodes share with one another, must be able to infer which action was taken.  In order to enable the system to be easily monitored, \emph{we further require this user observing no state to be able to learn an optimal action $\action_{\timeIdx}$ selected exclusively from the information shared during the time slot $\timeIdx$ during which the omniscient control action $\action_{\timeIdx}$ must be taken}.

For such a strategy, a key question is how much information must be shared in order to enable the optimal control action $\action_{\timeIdx}^* = \controlMap(\globState_{\timeIdx})$ the omniscient controller would have taken to be selected based on the shared information.  In other words, how much information must be shared in order to enable the system to simulate the omniscient controller in the sense that every node, including the one having access to no local state observations and only observing the shared control messages during time slot $\timeIdx$, can learn the action the  controller will take, thereby enabling the distributed system to obtain the same expected (discounted) long run reward as the omniscient system.

The answer to this question depends, of course, on the model for the way the information is exchanged, the control map $ \controlMap \in \candControlMapSet$, and particular characteristics of the transition kernels $\mdpTransDist_{\actiona }(\globStatei,\globStatej)$.  Clearly the problem of designing this communication has been transformed into one of distributed function computation, as each node observing a local state $\locState_{\nodeIdx,\timeIdx}$ must convey a message $\jmessage_{\nodeIdx,\timeIdx}$ during the time slot $\timeIdx$ such that $\action_{\timeIdx}^* = \controlMap(\globState_{\timeIdx})$ can be learned from the messages $\jmessage_{\nodeIdx,\timeIdx},\ \nodeIdx\in\nodeSet$.  Additionally, the capability to select any $\controlMap \in \candControlMapSet$ and still achieve the maximum reward enables this amount of information exchanged to be further minimized over  $\controlMap \in \candControlMapSet$.

\begin{example}[Downlink Wireless Resource Allocation, Continued]
\label{exmp:resourceallocation2}
The assumptions made above, and the associated problem of minimizing overhead, have special practical significance for the downlink wireless resource allocation setup of example \ref{exmp:resourceallocation1}.  Only the basestation has direct access to the buffer and observes the amount of information that has arrived destined for the various users, and only the users observe their downlink channel qualities, yet sufficient information must be exchanged for the system to make an informed decision regarding who to schedule on the downlink and how much information to send to them.  The omniscient controller having access to all of this state distributed throughout the network could make a series of decisions by solving the associated MDP.  At present, for instance in the LTE and WiMax standards, these decisions are made by the basestation, which requests and receives channel quality statistics from the users, then schedules the users and how much information to send to them, transmitting its decisions on the downlink \cite{Ku_CST_13}.  Additionally, it is desirable to minimize this amount of control information through an efficient design, as this type of control measurement and decision information, together with reference signals, has reached roughly a quarter to a third of the time frequency footprint in LTE and LTE advanced \cite{Ku_CST_13}.  Furthermore, it is essential that the messages exchanged during time slot $\timeIdx$ are all that must be overhead in order to learn the control decision action, as nodes come and go from the network, and it is essential that nodes that have just arrived in the current slot be able to determine what the control decisions were.  Finally, we note that it is evident from the problem description for example \ref{exmp:resourceallocation1}, that the various local states, which are the channel state at each user and the buffer state at the basestation, evolve according to independent Markov chains given the control actions.
\end{example}

\section{Minimal Coordination Communication Required for Distributed Simulation of the MDP}
\label{sec:min-communication-cost}
For general models with arbitrary dependence between observations, multiterminal information theory has yet to determine the minimum sum rate required for distributed function computation, however, these limits are known for a handful of special cases, including those where the local observations are independent.  In this independent case, the transition kernel and initial state distribution admit a factorization
\begin{equation}
\mdpTransDist_{\actiona }(\globStatei,\globStatej) = \prod_{\nodeIdx \in \nodeSet} \mdpTransDist_{\actiona}(\locStatei_{\nodeIdx},\locStatej_{\nodeIdx}),\ \ \mathbb{P}[\globState_{0} = \globStatei] = \prod_{\nodeIdx\in\nodeSet} \mathbb{P}[\locState_{\nodeIdx,0} = \locStatei_{\nodeIdx}].
\end{equation}
This factorization, in turn, implies that the local states evolve independently of one another, once an action has been specified, and as such, the quantities available to be encoded into messages at the nodes are independent of one another, so that $\mathbb{P}[\globState_{\timeIdx}=\globStatei] = \prod_{\nodeIdx \in \nodeSet} \mathbb{P}[\locState_{\nodeIdx,\timeIdx} = \locStatei_{\nodeIdx}] \ \forall \globStatei \in \globStateSet$.  The fundamental limits for this special case can be further subdivided based upon whether the messages $\jmessage_{\nodeIdx,\timeIdx}$ must all be sent in parallel and in a non-interactive manner, or if interaction between users over multiple rounds of communication during one time slot is allowed.  

\subsection{One Shot, Non-Interactive Distributed Simulation of the Omniscient MDP}\label{sec:noninteract}
In the non-interactive case, under the assumptions made regarding the monitoring node, the minimum sum-rate required for distributed function computation of a given control map $\controlMap \in \candControlMapSet$ with independent sources (in this case, the independent states) is given by the sum of the graph entropies of the characteristic graphs for each user \cite{FunctionComputing}\cite{FunctionCompression}.

The characteristic graph $\charGraph_{\nodeIdx}(\controlMap)$ for user $\nodeIdx$ has as its set of nodes $\charGraphNodes_{\nodeIdx} = \locStateSet_{\nodeIdx}$ the possible local states of user $\nodeIdx$.  An edge $\{\locStatei_{\nodeIdx},\locStatej_{\nodeIdx} \} \in \charGraphEdges_{\nodeIdx}$ exists in the characteristic graph if there are values $\locStatei_{\nodeIdx'} \in \locStateSet_{\nodeIdx'}, \nodeIdx' \in \nodeSet\setminus\{\nodeIdx\}$ such that $\mathbb{P}[\globState_{\timeIdx}= (\locStatei_{1},\ldots,\locStatei_{\nodeIdx-1},\locStatei_{\nodeIdx},\locStatei_{\nodeIdx+1},\ldots,\locStatei_{\numNodes})]>0$ and $\mathbb{P}[\globState_{\timeIdx}= (\locStatei_{1},\ldots,\locStatei_{\nodeIdx-1},\locStatej_{\nodeIdx},\locStatei_{\nodeIdx+1},\ldots,\locStatei_{\numNodes})]>0$ and $\controlMap(\locStatei_{1},\ldots,\locStatei_{\nodeIdx-1},\locStatei_{\nodeIdx},\locStatei_{\nodeIdx+1},\ldots,\locStatei_{\numNodes}) \neq \controlMap(\locStatei_{1},\ldots,\locStatei_{\nodeIdx-1},\locStatej_{\nodeIdx},\locStatei_{\nodeIdx+1},\ldots,\locStatei_{\numNodes})$.  

Since the local states are independent, the probability distribution will be positive on a product support $\locStateSet^{>0}_{\nodeIdx},\nodeIdx\in\nodeSet$, and there is no edge, i.e. $\{\locStatei_{\nodeIdx},\locStatej_{\nodeIdx} \} \notin \charGraphEdges_{\nodeIdx}$, if for all possible values of other local states $\locStatei_{\nodeIdx'} \in \locStateSet_{\nodeIdx'}^{>0}, \nodeIdx' \in \nodeSet\setminus\{\nodeIdx\}$, $\controlMap(\locStatei_{1},\ldots,\locStatei_{\nodeIdx-1},\locStatei_{\nodeIdx},\locStatei_{\nodeIdx+1},\ldots,\locState_{\numNodes}) = \controlMap(\locStatei_{1},\ldots,\locStatei_{\nodeIdx-1},\locStatej_{\nodeIdx},\locStatei_{\nodeIdx+1},\ldots,\locState_{\numNodes})$.  As such, there is a transitive property in the complement of the characteristic graph: namely if there is no edge $\{\locStatei_{\nodeIdx},\locStatej_{\nodeIdx} \}$ and no edge $\{\locStatej_{\nodeIdx},\locStatek_{\nodeIdx} \}$ in the characteristic graph, then there is also no edge  $\{\locStatei_{\nodeIdx},\locStatek_{\nodeIdx} \}$,  \emph{therefore the maximal independent sets of the characteristic graph do not overlap, and form a partition of the set of vertices of the graph}.  Owing to this transitivity property in the complement of the characteristic graphs \cite{FunctionComputing}\cite{RenBoyleKu2015}, each of these graph entropies is in fact the chromatic entropy
\begin{equation}
\entropy_{\charGraph_{\nodeIdx}(\controlMap)}(\locState_{\nodeIdx,\timeIdx}) = \min_{\jcoloring \in \allColorings(\charGraph_{\nodeIdx}(\controlMap))} \entropy(\jcoloring(\locState_{\nodeIdx,\timeIdx}))
\end{equation}
where $\allColorings(\charGraph_{\nodeIdx}(\controlMap))$ represent all colorings of the characteristic graph $\charGraph_{\nodeIdx}(\controlMap)$.  \emph{This minimum expected rate can be achieved within one bit by Huffman coding the coloring of the characteristic graph achieving the minimum entropy, which then can be achieved by assigning different colors to its different maximal independent sets.}
 Selecting an omniscient control map $\controlMap \in \candControlMapSet$ requiring the minimum rate, then gives the minimum non-interactive rate of
\begin{equation}\label{eq:noIntLimit}
\minRateNoInt = \min_{\controlMap \in \candControlMapSet} \sum_{\nodeIdx\in\nodeSet} \min_{\jcoloring \in \allColorings(\charGraph_{\nodeIdx}(\controlMap))} \entropy(\jcoloring(\locState_{\nodeIdx,\timeIdx}))
\end{equation}
Note further than when searching minimum entropy colorings of the graph to calculate the chromatic entropy, it suffices to consider exclusively the \emph{greedy}-colorings \cite{MinimumEntropyColoring} obtained by iteratively removing maximal independent sets.

The following two examples describe the control rate required under this form of one-shot, non-interactive sharing of quantized local states, for two particular distributed MDPs.

\begin{example}[Minimum control information for $\arg\max$, Non-Interactive]
\label{exmp:argmaxNonInteractive}

	Let's assume in example \ref{exmp:resourceallocation2} that the buffer size is infinite, and each user has infinitely many backlogged packets destined for it in the buffer. Then the control decision is made regarding only the users' channel qualities, and the objective is to let the basestation learn the control decision of which user should occupy the resource block after observing all the messages sent from the users.  
	
	
	Let the local states $\locState_1,\ldots,\locState_N$ be independent and identically distributed downlink channel qualities from a known distribution on a discrete support set $\locStateSet$, if the basestation wishes to maximize the system throughput, the control decision becomes finding one of the users with the best channel quality, i.e.
\begin{equation}
\controlMap(\locState_1,\ldots,\locState_N) \in \underset{\nodeIdx \in \nodeSet}{\arg\max}\ \locState_{\nodeIdx} 
\end{equation}
For this problem, it is shown in \cite{RenBoyleKu2015} that the characteristic graphs $\charGraph_{1}(\controlMap),\ldots,\charGraph_{\numNodes}(\controlMap)$ obey the properties that if $\{\locStatei,\locStatej \} \notin \charGraphEdges_{\nodeIdx}$ then $\{\locStatei,\locStatek \} \in \charGraphEdges_{\nodeIdx}, \{\locStatej,\locStatek \} \in \charGraphEdges_{\nodeIdx}, \forall \locStatek \in \locStateSet$,
and that if $\{\locStatei,\locStatej \} \notin \charGraphEdges_{\nodeIdx}$ then $\{\locStatei,\locStatej \} \in \charGraphEdges_{\nodeIdx'}, \forall \nodeIdx' \in \nodeSet \setminus \{\nodeIdx\}$. It is also shown in \cite{RenBoyleKu2015} that the minimum information required to determine the control action can be computed as in (\ref{eq:noIntLimit}), and at most $2$ bits can be saved relative to the scheme in which the users simply send their un-coded channel qualities to the basestation.

\end{example}

\begin{example}[Rate Required for Simulating an Omniscient Wireless Resource Controller with No Interaction]
\label{exmp:one-shot}
Return to the case of a finite buffer size without any backlogged packets in example \ref{exmp:resourceallocation2}, and assume that the channel qualities
where $\locState_{\nodeIdx,\timeIdx}$ are independently uniformly distributed on the support $\{0,1,2,3\}\ \forall \nodeIdx \in \nodeSet, \timeIdx \in \mathbb{N}$.  In addition, let the amount of additional traffic that arrives destined for each user be independent across users and time, and be distributed on the support $\mathcal{X} = \{0,1,2\}$ with probabilities $\{\nicefrac{1}{2},\nicefrac{1}{3},\nicefrac{1}{6}\}$.   Additionally, let the packet dropping function operate according to Algorithm \ref{algorithm:packagedrop}, in a manner consistent with a total buffer size of $BU_{max}=3$.   Let the controller aim to maximize the throughput reward (\ref{eq:throughput}).   If $\numNodes = 2$, which means there are $2$ users and $1$ basestation in the system, the optimal control decisions by solving the MDP problem of (\ref{eq:optValue}) and (\ref{eq:cand-optControl}) with discounting factor $\discountFactor = 0.9$ will give a maximal total discounted reward of $9.249$ if the system starts from the all $0$ initial state $\globState_0 = (\locState_{1,0},\locState_{2,0},\locState_{3,0}) = (0,0,(0,0))$. Meanwhile, the expected amount of system throughput per time-slot will be $1.076$ and the expected amount of data dropped per time-slot will be $0.257$.  Calculating the characteristic graphs and determining the Huffmann codes associated with the minimum entropy colorings, we find that the associated optimal control decision can be learned via a quantization of local states (the channel states at each of the two users and the buffer size at the basestation), with a minimum non-interactive rate of  $3.5175$ bits. The encoder mappings with respect to the minimum rate are given as:
\begin{equation}
\locStateEncoder_1(\locState_1) = \begin{dcases*}
2 & if $\locState_1 = 0$\\
1 & if $\locState_1 \in \{1,2,3\}$\\
\end{dcases*}
\end{equation}
\begin{equation}
\locStateEncoder_2(\locState_2) = \begin{dcases*}
1 & if $\locState_2 \in \{0,1\}$\\
2 & if $\locState_2 \in \{2,3\}$\\
\end{dcases*}
\end{equation}
for the users, and
\begin{equation}
\locStateEncoder_3(\locState_3) = \begin{dcases*}
1 & if $\locState_3 \in \{(1,0),(2,0),(3,0),(1,1),(2,1)\}$\\
2 & if $\locState_3 \in \{(0,1),(0,2),(0,3),(0,0)\}$\\
3 & if $\locState_3 \in \{1,2\}$\\
\end{dcases*}
\end{equation}
for the basestation. A control mapping $\controlMap': \locStateEncoderVec(\globStateSet) \rightarrow \actionSet$ can be decided deterministically with the given optimal control $\controlMap$ and the encoders $\locStateEncoderVec$, i.e.
$
\controlMap'(\locStateEncoder_1(\locState_1)=1,\locStateEncoder_2(\locState_2)=2, \locStateEncoder_3(\locState_3)=3) = 2
$
and
$
\controlMap'(\locStateEncoder_1(\locState_1)=1,\locStateEncoder_2(\locState_2)=1, \locStateEncoder_3(\locState_3)=3) = 1
$.
\begin{algorithm}[h]
\DontPrintSemicolon
\KwResult{Update next round buffer status $\locState_{\numNodes+1,\timeIdx+1}= (B_{1,\timeIdx+1},\ldots,B_{\numNodes,\timeIdx+1})$}
Input:\ current buffer status $\locState_{\numNodes+1,\timeIdx} = (B_{1,\timeIdx},\ldots,B_{\numNodes,\timeIdx})$, buffer size $BU_{max}$, and new arriving packages $\boldsymbol{X}_{\timeIdx} = (X_{1,\timeIdx},\ldots,X_{\numNodes,\timeIdx})$\;
$\locState_{\numNodes+1,\timeIdx+1} = \locState_{\numNodes+1,\timeIdx}$ \;
$\text{Remaining} = BU_{max} - \sum_{\nodeIdx=1}^{\numNodes} B_{\nodeIdx,t}$ \;
$\text{New} = \sum_{\nodeIdx=1}^{\numNodes} X_{\nodeIdx,\timeIdx+1}$ \;
$\nodeIdx = 1$\; 
\While{$\text{Remaining} > 0$ \& $\text{New} > 0$}{
\If{$X_{\nodeIdx,\timeIdx+1} > 0$}{
$B_{\nodeIdx,\timeIdx+1}= B_{\nodeIdx,\timeIdx+1} + 1$\;
$X_{\nodeIdx,\timeIdx+1} = X_{\nodeIdx,\timeIdx+1} - 1$\;
$\text{Remaining} = \text{Remaining} - 1$\;
$\text{New} = \text{New} - 1$\;
}
$\nodeIdx = (\nodeIdx + 1) \mod \numNodes$ \;
}
Output $\locState_{\numNodes+1,\timeIdx+1}$\;
\caption{The packet dropping process when the buffer is full \label{algorithm:packagedrop}}
\end{algorithm}

\end{example}

\subsection{Interactive, Collocated Network, Simulation of the Omniscient MDP}\label{sec:interact}
In the interactive communication case, a natural lower bound on the rate can be obtained via a collocated network messaging model.  In this model, communication happens over multiple rounds, in which users consecutively take turns sending a message which is overheard by all of the other users.  In particular, in round $\roundIdx$, the node with index $\nodeIdxFunc(\roundIdx) = (\roundIdx-1)\textrm{mod} | \nodeSet| + 1$ sends a message $\jmessage_{\timeIdx}^{(\roundIdx)}$ based on its observation $\locState_{\nodeIdxFunc(\roundIdx),\timeIdx}$ and all of the messages $\jmessage_{\timeIdx}^{(1)},\ldots,\jmessage_{\timeIdx}^{(\roundIdx-1)}$ sent in the previous rounds up until this time.  After $\numRounds$ rounds the communication finishes and the optimum omniscient control action $\action_{\timeIdx}^* = \controlMap(\globState_{\timeIdx})$ must be completely determined from $\jmessage_{\timeIdx}^{(1)},\ldots,\jmessage_{\timeIdx}^{(\numRounds)}$.  Ma and Ishwar \cite{InteractiveCollocated} have shown that the minimum sum-rate, over all block codes, that can be obtained by such a strategy is lower bounded by the solution to following repeated convex geometric calculation.  The solution is written with respect to the rate reduction functional, which maps the coordinates for the marginal probability distributions $\probDistrVec_{\nodeIdx} \in \probDistrSet(\locStateSet_{\nodeIdx})$ for each of the local observations $\nodeIdx \in \nodeSet$ to a conditional entropy,
\begin{equation}
\begin{aligned}
\rateReduc_{\roundIdx}: &\prod_{\nodeIdx\in\nodeSet} \probDistrSet( \locStateSet_{\nodeIdx}) \rightarrow \Reals, \quad \rateReduc_{\roundIdx}(\probDistrVec ) := \entropy( \globState |\jmessage^{(1)},\ldots, \jmessage^{(\roundIdx)})\\
 &\textrm{if the marginal distribution of } \ \globState \ \textrm{is given by}\ \probDistrVec.\\
\end{aligned}
\end{equation}
The rate required if the function is to be computed after $\numRounds $ rounds of communication is then expressed as
\begin{equation}\label{eq:intRateLimit}
\minRateWI_{\numRounds} = \entropy(\globState_{\timeIdx}) - \rateReduc_{\roundIdx}(\probDistr_{\globState_{\timeIdx}}).
\end{equation}
i.e. by evaluating the rate reduction functional at the marginal probability distribution $\probDistr_{\globState_{\roundIdx}}$.  The rate reduction functional, in turn, is found via the following iterative convex program
\begin{equation}\label{convex_geomertic}
\rateReduc_{0}(\probDistrVec) = \left\{ \begin{array}{cc}  \entropy(\probDistrVec ) & \exists \controlMap \in \candControlMapSet \ s.t.\  \controlMap \ \textrm{is constant on }\ \textrm{supp}(\probDistrVec) \\ -\infty & \textrm{ otherwise} \end{array}  \right.
\end{equation}
\begin{equation}
\begin{aligned}\label{eq:convProg}
\rateReduc_{\roundIdx}&(\probDistr_{1},\ldots,\probDistr_{\nodeIdxFunc(\roundIdx)-1},\cdot ,\probDistr_{\nodeIdxFunc(\roundIdx)+1},\ldots,\probDistr_{\numNodes}) =\\ 
&\upperConcave(\rateReduc_{\roundIdx - 1}(\probDistr_{1},\ldots,\probDistr_{\nodeIdxFunc(\roundIdx)-1},\cdot ,\probDistr_{\nodeIdxFunc(\roundIdx)+1},\ldots,\probDistr_{\numNodes})) \\
\end{aligned}
\end{equation}
Here for each fixed context $\probDistr_{\nodeIdx} \in \probDistrSet(\locStateSet_{\nodeIdx}),\ \nodeIdx \in \nodeSet \setminus \{ \nodeIdxFunc(\roundIdx) \}$,  the operator $\upperConcave$ is computing the upper concave envelope of the function $\rateReduc_{\roundIdx - 1}(\probDistr_{1},\ldots,\probDistr_{\nodeIdxFunc(\roundIdx)-1},\cdot ,\probDistr_{\nodeIdxFunc(\roundIdx)+1},\ldots,\probDistr_{\numNodes})$, i.e. viewing this restriction as a function from $\probDistrSet(\locStateSet_{\nodeIdx}) \rightarrow \Reals$.

Note that for the problem at hand, this lower bound may not be achievable, as scalar quantization and coding strategies are required by the assumptions we have made in our distributed MDP setup, while this lower bound may in general only be achieved with a limit of vector quantization schemes.  In particular, only scalar quantization is available in the problem under consideration because no repeated observations are available for use in a larger block-length as we have added the constraint that the user overhearing exclusively all of the messages during time slot $\timeIdx$ must be able to learn $\action_{\timeIdx}^* = \controlMap(\globState_{\timeIdx})$.   Nonetheless, as we demonstrate in the following example, often the best scalar quantization based interaction schemes still yield a rate which is very close to this fundamental limit.

\begin{figure}[t]
	\centering
\includegraphics[width=.97\columnwidth]{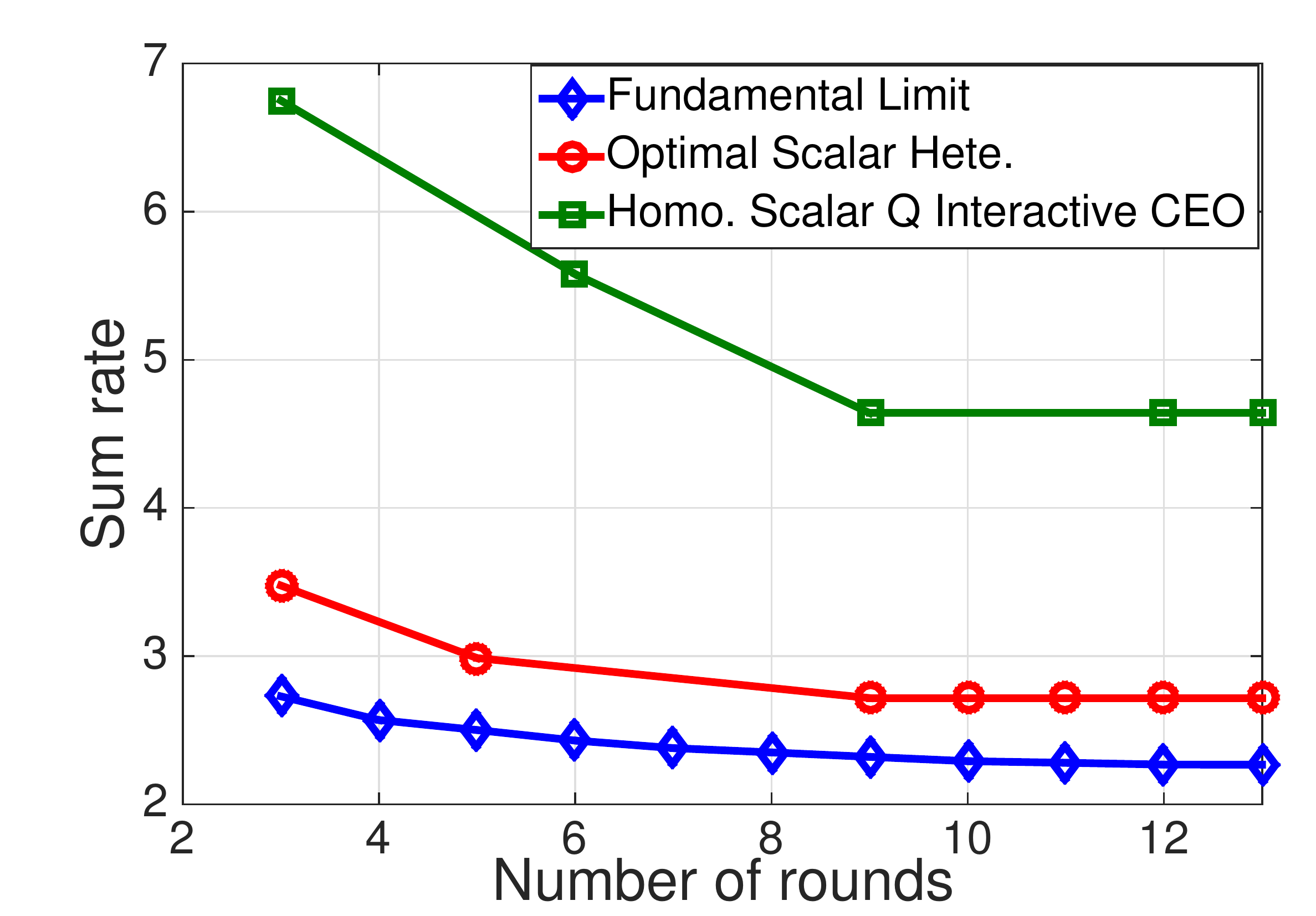}
	\caption{Rate required to select a user attaining the max with 3 users each observing an independent RV with support $\{1,2,3,4\} $ when users take turns sending messages one at a time \cite{SolmazDCC2016,SolmazISIT2016}.  The fundamental limit on the rate given by (\ref{eq:intRateLimit}) is compared to that obtained by the best scalar quantizer to use in the collocated network interactive scheme (Optimal Scalar. Hete. Q).   This is substantially less than the rate required if all three users must send each message in parallel, then hear collectively all of the sent messages, then send each message in parallel again, etc, which is the interactive model employed in \cite{Boyle_RDfuncComp}.}
	\label{fig:interactiveRateReq}
\end{figure}

\begin{example}[Minimum control information for $\arg\max$, Interactive]
Consider the infinite packet buffer backlog and throughput maximization variant of the wireless resource allocation as described in Example  \ref{exmp:argmaxNonInteractive}, with the added ability that the users and basestation can all interact with one another while sending their messages.  In particular, the users can each take turns sending messages, one at a time, over $\numRounds$ rounds, such that all the other participants, including the basestation which sends no messages, overhear each message.  The goal is to enable anyone who overhears all of these messages to learn the index of at least one user whose channel quality is the same as the maximum channel quality over all of the users.  The curve labelled Fundamental limit in Fig. \ref{fig:interactiveRateReq} calculates the fundamental limit (\ref{eq:convProg}) lower bounding the total number of bits must be exchanged in order to perform this calculation for the case of $\numNodes=3$ and for channel qualities uniformly distributed over the set $\{1,2,3,4\}$.  As explained above, this fundamental limit is in general only achievable with vector quantization schemes, while the problem setup at hand demands that scalar quantization schemes must be used.  Additionally, a second curve in Fig. \ref{fig:interactiveRateReq}, labelled ``optimal scalar Hete. Q''  gives the rate required by the best possible scalar quantization scheme, followed by Huffman coding, for this problem, and this is seen to be quite close to the vector quantization limit.  This curve was found via exhaustive search over all scalar quantization schemes.  In addition to presenting this problem in detail and describing these curves,\cite{SolmazISIT2016} also considers reduced complexity  restricted to smaller quantizer search spaces.  Finally, the curve ``Homo. scalar Q interactive'' indicates the rate required if all of the users must send their messages in parallel, then, after all of these messages are received, can send another series in parallel and so on, which is the interactivity model considered in \cite{Boyle_RDfuncComp}.  While this does count as a form of interaction, requiring the users to send their messages in parallel substantially increases the rate required.
\end{example}

\section{Incorporating Communication Cost into the Reward Function}\label{sec:iterative}
The previous discussion has assumed that the reward function is completely given, but in many problems, the cost of communicating over the network may subtract from the reward of the decisions.    In this manner, it may be desirable to design the MDP to consider this cost explicitly by incorporating it as a weighted term into the reward function.  In particular, suppose that the total number of bits communicated by the partial state sharing scheme in the messages $\jmessage_{\nodeIdx, \timeIdx}, \nodeIdx \in \nodeSet$ when the state vector is $\globState_{\timeIdx} = \globStatei$ is denoted by $|\locStateEncoderVec(\globStatei)|$, then we can form an augmented reward function
\begin{equation}\label{eq:augmented-reward}
\rewardFunc_{\actiona}'(\globStatei,\globStatej) = \rewardFunc_{\actiona}(\globStatei,\globStatej) - \controlRateLagrMul |\locStateEncoderVec(\globStatei) |
\end{equation}
including the communication cost reflecting the number of bits transmitted in order to enable the system learn the action.  The goal then shifts to solving the optimization problem
\begin{equation}\label{eq:constrMDP}
\max_{(\controlMap,\locStateEncoderVec) \in \feasSet}  \sum_{\timeIdx \in \mathbb{N} } \discountFactor^{\timeIdx} \left( \mathbb{E}\left[\rewardFunc_{\controlMap(\globState_{\timeIdx})}(\globState_{\timeIdx},\globState_{\timeIdx+1}) - \controlRateLagrMul  |\locStateEncoderVec(\globState_{\timeIdx}) | \right] \right)
\end{equation}
where the constraint set is defined as
\begin{equation}
\begin{aligned}
\feasSet :=& \left\{ (\controlMap,\locStateEncoderVec) \left| \forall \globStatei,\globStatei' \in \globStateSet \ \textrm{such that}\ \locStateEncoderVec(\globStatei) = \locStateEncoderVec(\globStatei'),\right.\right.\\ 
&\left.\left.\textrm{we have also that}\ \controlMap(\globStatei) = \controlMap(\globStatei'). \right. \right\}\\
\end{aligned}
\end{equation}
or equivalently, the set of $\controlMap,\locStateEncoderVec$ such that $\controlMap$ can be rewritten as a function of exclusively $\locStateEncoderVec$,
\begin{equation}
\begin{aligned}
\feasSet := \bigl\{ (\controlMap,\locStateEncoderVec)\bigl | \locStateEncoderVec: \globStateSet \rightarrow \binStrings,\ &\exists \controlMap':\locStateEncoderVec(\globStateSet) \rightarrow \actionSet,\\ 
&\controlMap(\globStatei) = \controlMap'(\locStateEncoderVec(\globStatei))\ 
\forall \globStatei \in \globStateSet \bigr\}\\
\end{aligned}
\end{equation}
Observe that while the observation of the encodings $\locStateEncoderVec(\globStatei)$ form effectively an observation for a partial observed Markov decision process (POMDP) \cite{VikramPOMDP}\cite{MurphyPOMDP}, the requirement we have made that we are able to determine the controller's (i.e. running the full state knowledge MDP) action decisions from exclusively the observation during the same time step implies a different problem structure from a POMDP as the memory of past observations or action decisions for determining the state distribution must be neglected.  

Select some map $\stateOrderer:\globStateSet \rightarrow \{1,\ldots,|\globStateSet|\}$ and, for any given controller $\controlMap : \globStateSet \rightarrow \actionSet$, define the $|\globStateSet| \times | \globStateSet|$ transition matrix $\transMatrix(\controlMap) $ whose $\stateOrderer(\globStatei),\stateOrderer(\globStatej)$th element is
\begin{equation}
\begin{aligned}
[\transMatrix(\controlMap)&]_{\stateOrderer(\globStatei),\stateOrderer(\globStatej)} = \mdpTransDist_{\controlMap(\globStatei)}(\globStatei,\globStatej), \\
&\textrm{and define the row vector} \quad [\initStateDistrV]_{\stateOrderer(\globStatei)} = \mathbb{P}[\globState_{0}=\globStatei].
\end{aligned}
\end{equation}
The objective function in the optimization can then be rewritten as
\begin{equation}
\begin{aligned}
&\sum_{\timeIdx \in \mathbb{N} } \discountFactor^{\timeIdx} \left( \mathbb{E}\left[\rewardFunc_{\controlMap(\globState_{\timeIdx})}(\globState_{\timeIdx},\globState_{\timeIdx+1}) - \controlRateLagrMul  |\locStateEncoderVec(\globState_{\timeIdx}) | \right] \right)\\ 
&=\sum_{\timeIdx \in \mathbb{N} } \discountFactor^{\timeIdx}  \sum_{\globStatei,\globStatej\in\globStateSet} \mathbb{P}[\globState_{\timeIdx}=\globStatei,\globState_{\timeIdx+1}=\globStatej] \left( \rewardFunc_{\controlMap(\globStatei)}(\globStatei,\globStatej) - \controlRateLagrMul  |\locStateEncoderVec(\globStatei) |  \right) \\
&= \sum_{\timeIdx \in \mathbb{N} } \discountFactor^{\timeIdx}  \sum_{\globStatei,\globStatej\in\globStateSet} [\initStateDistrV \transMatrix(\controlMap)^{\timeIdx}]_{\stateOrderer(\globStatei)} \mdpTransDist_{\controlMap(\globStatei)}(\globStatei,\globStatej) \left( \rewardFunc_{\controlMap(\globStatei)}(\globStatei,\globStatej) - \controlRateLagrMul  |\locStateEncoderVec(\globStatei) |  \right) \\
&= \sum_{\globStatei,\globStatej\in\globStateSet} \left[\initStateDistrV \sum_{\timeIdx \in \mathbb{N} } \discountFactor^{\timeIdx} \transMatrix(\controlMap)^{\timeIdx}\right]_{\stateOrderer(\globStatei)} \mdpTransDist_{\controlMap(\globStatei)}(\globStatei,\globStatej) \left( \rewardFunc_{\controlMap(\globStatei)}(\globStatei,\globStatej) - \controlRateLagrMul  |\locStateEncoderVec(\globStatei) |  \right) \\
&= \sum_{\globStatei,\globStatej\in\globStateSet} \left[\initStateDistrV (\identityMatrix - \discountFactor \transMatrix(\controlMap))^{-1} \right]_{\stateOrderer(\globStatei)} \mdpTransDist_{\controlMap(\globStatei)}(\globStatei,\globStatej) \left( \rewardFunc_{\controlMap(\globStatei)}(\globStatei,\globStatej) - \controlRateLagrMul  |\locStateEncoderVec(\globStatei) |  \right) 
\end{aligned}
\end{equation}
The presence of the constraint that $(\controlMap,\locStateEncoderVec)\in\feasSet$ makes the joint optimization problem (\ref{eq:constrMDP}) a substantially more difficult combinatorial optimization problem than an ordinary MDP.  For small problems, the set of control maps $\controlMap:\globStateSet \rightarrow \actionSet$ can be enumerated, and for each such control map, the component
\begin{equation}
\max_{\locStateEncoderVec| (\controlMap,\locStateEncoderVec)\in\feasSet}  - \controlRateLagrMul  \sum_{\globStatei\in\globStateSet} \left[\initStateDistrV (\identityMatrix - \discountFactor \transMatrix(\controlMap))^{-1} \right]_{\stateOrderer(\globStatei)}|\locStateEncoderVec(\globStatei) |
\end{equation}
of the expected reward associated with the minimum control information overhead can be calculated using the results in section \ref{sec:noninteract} for a non-interactive messaging scheme, while if interactive communications are enabled, then the results in section \ref{sec:interact} can be used.   In both cases, the probability distribution for $\globState$ is selected as being multiplicatively proportional to $ \left[\initStateDistrV (\identityMatrix - \discountFactor \transMatrix(\controlMap))^{-1} \right]_{\stateOrderer(\globStatei)}$ to ensure that the reward will be maximized by the encoding.  After determining in this manner the expected discounted reward $\max_{\locStateEncoderVec| (\controlMap,\locStateEncoderVec)\in\feasSet}\sum_{\timeIdx \in \mathbb{N} } \discountFactor^{\timeIdx} \left( \mathbb{E}\left[\rewardFunc_{\controlMap(\globState_{\timeIdx})}(\globState_{\timeIdx},\globState_{\timeIdx+1}) - \controlRateLagrMul  |\locStateEncoderVec(\globState_{\timeIdx}) | \right] \right) $, maximized over all encoding schemes, for each control map $\controlMap$, the control map yielding the expected maximum reward for the particular $\controlRateLagrMul$ can be selected.  Furthermore, a tradeoff between the control overhead and the expected reward can be traced by varying $\controlRateLagrMul$ in this optimization.

\begin{figure*}
	\centering
	\subfloat[]{\includegraphics[width=0.5\textwidth]{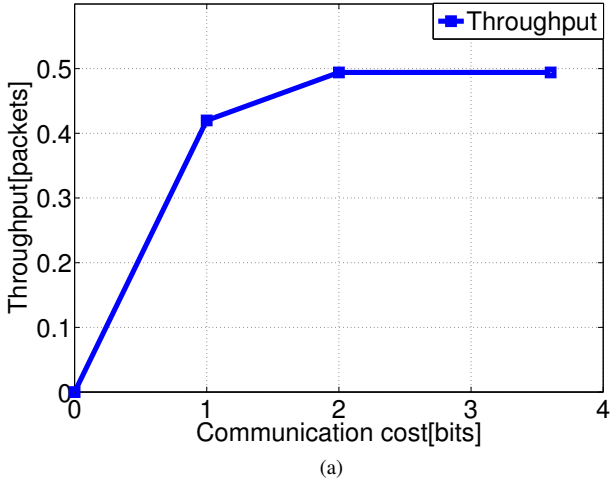}
	\label{fig:search-all-throughput}}
	\subfloat[]{\includegraphics[width=0.5\textwidth]{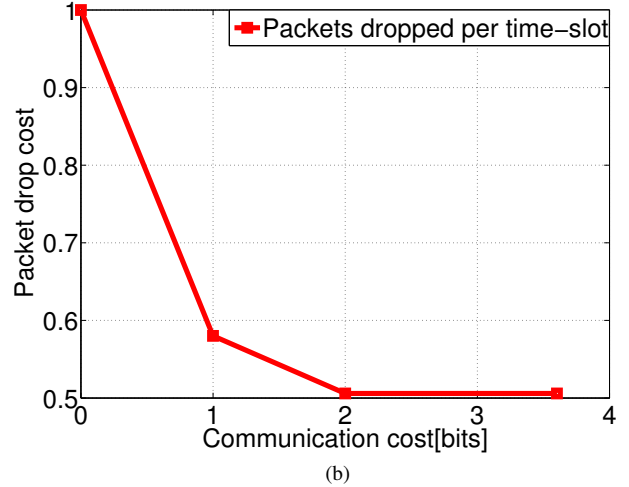}
	\label{fig:search-all-drop}}
	\caption[]{(a) The control overhead versus the expected throughput tradeoff, and (b) The control overhead versus the expected packet dropping cost tradeoff with $\controlRateLagrMul \in [0,0.36]$ in the augmented reward function in (\ref{eq:augmented-reward})}
	\label{fig:searchall-tradeoff}
\end{figure*}
\begin{example}[Overhead Performance Tradeoff for Wireless Resource Allocation, Tiny Model]\label{exmp:tinyexact}
Consider again the setup of distributed wireless resource allocation described in example \ref{exmp:resourceallocation1}, with $\numNodes = 2$ mobile users each observing their local channel quality $\locState_{\nodeIdx,\timeIdx}$ at each time instant $\timeIdx$,
where $\locState_{\nodeIdx,\timeIdx}$ is uniformly distributed on the support $\{0,1\} \forall \nodeIdx \in \nodeSet, \timeIdx \in \{1,\ldots\}$.  Additionally, let the amount of additional traffic that arrives destined for each user $X_{\nodeIdx,\timeIdx}$ be uniformly distributed on the support $\mathcal{X} = \{0,1\}$, and let the buffer size limit at the basestation be $BU_{max} = 2$, with the packet dropping process  described as in Example \ref{exmp:one-shot}.   The system is started in the all $0$ initial state, in which the buffer is empty and the channel qualities are all $0$.  Let (\ref{eq:augmented-reward}) be the transition reward function where the term $\rewardFunc_{\actiona}(\globStatei,\globStatej)$ represents the system throughput by choosing action $\actiona$ at state $\globStatei$ as in (\ref{eq:throughput}), and consider the one-shot, non-interactive, control information sharing model in which each node (i.e. each user and the base station) sends a quantized representation of their state to everyone else, enabling them all to learn the control action directly from these messages.  Finding the optimum solution to (\ref{eq:constrMDP}) via exhaustive search over all control mappings for different $\controlRateLagrMul$ enables the control overhead versus throughput tradeoff and the control overhead versus the packet dropping cost tradeoff plotted in Fig. \ref{fig:search-all-throughput} and \ref{fig:search-all-drop} to be traced out. We observe from Fig. \ref{fig:searchall-tradeoff} that, at least $2$ bits of control overheads are required to guarantee the expected system throughput achieves the limit.
%
\end{example}

\subsection{Finding Candidate Quantizations through Alternating Optimization}\label{sec:alg}
However, in many problems, the sort of exhaustive search approach to solving the combinatorial optimization (\ref{eq:constrMDP}) just described is nowhere near computationally feasible, as the number of possible control maps to search over are $|\actionSet|^{|\globStateSet|}$.  In this case, an alternating optimization approach yields a lower complexity search method that can be well suited to finding candidate solutions to this optimization problem.

A reasonable goal for such an alternating optimization method is to alternate between optimizing the control map, then optimizing the quantizer.  Let the iteration index in this algorithm be $\iterIdx$, and the control map and local state encoders at iteration $\iterIdx$ be denoted by $\controlMap_{\iterIdx}$ and $\locStateEncoderVec_{\iterIdx}$ respectively.   At a given iteration in the algorithm, the control map minimizing the augmented value function among all control maps that can be determined from the present encoding $\locStateEncoderVec_{\iterIdx}$ could be selected by solving
\begin{equation}\label{eq:controlUpdate}
\controlMap_{\iterIdx+1} = \underset{\controlMap | (\controlMap,\locStateEncoderVec_{\iterIdx}) \in \feasSet}{\arg \max}\sum_{\timeIdx \in \mathbb{N} } \discountFactor^{\timeIdx} \left( \mathbb{E}\left[\rewardFunc_{\controlMap(\globState_{\timeIdx})}(\globState_{\timeIdx},\globState_{\timeIdx+1}) - \controlRateLagrMul  |\locStateEncoderVec_{\iterIdx}(\globState_{\timeIdx}) | \right] \right)
\end{equation}
Next, the local encodings $\locStateEncoderVec_{\iterIdx+1}$ which achieve the minimum expected sum rate while enabling distributed computation of the new control map $\controlMap_{\iterIdx+1}$ are selected
\begin{equation}\label{eq:quantizationUpdate}
\begin{aligned}
&\locStateEncoderVec_{\iterIdx+1} =\\
& \underset{\locStateEncoderVec | (\controlMap_{\iterIdx+1},\locStateEncoderVec) \in \feasSet}{\arg \max} \sum_{\timeIdx \in \mathbb{N} } \discountFactor^{\timeIdx} \left( \mathbb{E}\left[\rewardFunc_{\controlMap^{(\iterIdx+1)}(\globState_{\timeIdx})}(\globState_{\timeIdx},\globState_{\timeIdx+1}) - \controlRateLagrMul  |\locStateEncoderVec(\globState_{\timeIdx}) | \right] \right)\\
\end{aligned}
\end{equation}
As this admits a form of an alternating maximization, the sequence of expected values will be monotone increasing.  As this sequence is bounded above via the global optimum, this sequence must converge to a limit.  In general this limit may or may not be the global optimum, as all that can be guaranteed is that this limit is associated with a Nash equilibrium.  In particular, the limit of this iteration has the property that in no unilateral change individually in the control map $\controlMap$ or the quantizer $\locStateEncoderVec$ can yield a higher expected reward, although it may be possible to modify them both together and achieve a higher reward.

To solve (\ref{eq:quantizationUpdate}), if the non-interactive communications structure is used, the results in section \ref{sec:noninteract} and equation (\ref{eq:noIntLimit}) can be used as a fairly tight and close bound with associated close achievability scheme (within one bit), while if interactive communications are enabled, then the results in section \ref{sec:interact} and equations (\ref{eq:intRateLimit},\ref{eq:convProg}) can be used as a bound.

Solving (\ref{eq:controlUpdate}) however, can be quite complicated, as the direct search solution to the combinatorial optimization has complexity $|\actionSet|^{|\globStateSet|}$.  To simplify matters, the control map update (\ref{eq:controlUpdate}) can itself be attacked with an alternating optimization which has an overall iteration update complexity proportional to $|\locStateEncoderVec(\globStateSet)| |\actionSet|$.  In its simplest form, this alternating minimization cycles through the different possible quantizations $\possibleMess \in \locStateEncoderVec(\globStateSet)$, updating the associated $\controlMap'(\possibleMess)=\controlMap(\globStatei), \ \globStatei \in \locStateEncoderVec^{-1}(\possibleMess)$ in an order determined by a selected bijection $\messOrderMap: \locStateEncoderVec(\globStateSet) \rightarrow \{0,1,\ldots, | \locStateEncoderVec(\globStateSet)|-1\} $ according to
\begin{equation}\label{eq:ord1}
\possibleMess_{\iterIdx,\iterIdxb} = \messOrderMap^{-1}(\iterIdxb\ \textrm{mod}\ | \locStateEncoderVec_{\iterIdx} (\globStateSet) |)  
\end{equation}
\begin{equation}\label{eq:ord2}
\begin{aligned}
&\controlMap_{\iterIdx,\iterIdxb} = \mathop{\arg\max}_{ \controlMap \in \feasSet(\possibleMess_{\iterIdx,\iterIdxb} ,\controlMap_{\iterIdx,\iterIdxb-1},\locStateEncoderVec_{\iterIdx})  } \sum_{\globStatei\in\locStateEncoderVec_{\iterIdx}^{-1}(\possibleMess_{\iterIdx,\iterIdxb})} \sum_{\globStatej \in \globStateSet} \\
&\left[\initStateDistrV (\identityMatrix - \discountFactor \transMatrix(\controlMap))^{-1} \right]_{\stateOrderer(\globStatei)} \mdpTransDist_{\controlMap(\globStatei)}(\globStatei,\globStatej) \left( \rewardFunc_{\controlMap(\globStatei)}(\globStatei,\globStatej) - \controlRateLagrMul  |\locStateEncoderVec(\globStatei) |  \right) \\
\end{aligned}
\end{equation}
\begin{equation}\label{eq:ord3} 
\controlMap_{\iterIdx+1} = \lim_{\iterIdxb \rightarrow \infty} \controlMap_{\iterIdx,\iterIdxb},\ \quad \controlMap_{\iterIdx,0} := \controlMap_{\iterIdx}. 
\end{equation}
wherein
\begin{equation}
\begin{aligned}
\feasSet(&\possibleMess_{\iterIdx,\iterIdxb} ,\controlMap_{\iterIdx,\iterIdxb-1},\locStateEncoderVec_{\iterIdx}) := \\
&\left\{  \controlMap: \globStateSet \rightarrow \actionSet \left| \begin{array}{c} \exists \controlMap':\locStateEncoderVec(\globStateSet) \rightarrow \actionSet,\ \controlMap = \controlMap' \circ \locStateEncoderVec_{\iterIdx},\\
\controlMap'(\possibleMess_{\iterIdx,\iterIdxb}) \in \actionSet, \controlMap'(\possibleMess) = \controlMap'_{\iterIdx,\iterIdxb-1}\\(\possibleMess)\ \forall \possibleMess \in  \locStateEncoderVec_{\iterIdx} (\globStateSet)  \setminus \{  \possibleMess_{\iterIdx,\iterIdxb}\}. \end{array} \right. \right\}.\\
\end{aligned}
\end{equation}

Alternatively, a more greedy form of the alternating optimization can be selected, which replaces (\ref{eq:ord1}) with
\begin{equation}\label{eq:greedy1}
\begin{aligned}
&\possibleMess_{\iterIdx,\iterIdxb} = \underset{\possibleMess' \in \locStateEncoderVec_{\iterIdx}(\globStateSet)}{\arg\max}\  \underset{\controlMap \in\feasSet(\possibleMess' ,\controlMap_{\iterIdx,\iterIdxb-1},\locStateEncoderVec_{\iterIdx})}{\max} \sum_{\globStatei\in\locStateEncoderVec_{\iterIdx}^{-1}(\possibleMess')} \sum_{\globStatej \in \globStateSet}\\
 &\left[\initStateDistrV (\identityMatrix - \discountFactor \transMatrix(\controlMap))^{-1} \right]_{\stateOrderer(\globStatei)} \mdpTransDist_{\controlMap(\globStatei)}(\globStatei,\globStatej) \left( \rewardFunc_{\controlMap(\globStatei)}(\globStatei,\globStatej) - \controlRateLagrMul  |\locStateEncoderVec(\globStatei) |  \right)
\end{aligned}
\end{equation}

Putting these pieces together, the overall low complexity alternating optimization algorithm to find candidate solutions to (\ref{eq:constrMDP}) consists of (\ref{eq:ord1}) or (\ref{eq:greedy1}) and (\ref{eq:ord2}),(\ref{eq:ord3}),and (\ref{eq:quantizationUpdate}).  As this algorithm is an alternating optimization, with the individual dimensions in the optimization being $\locStateEncoderVec$, and each of the $\controlMap'(\possibleMess), \possibleMess \in \locStateEncoderVec(\globStateSet)$, the sequence of expected rewards achieved by each update is monotone increasing.  As this sequence of expected rewards is bounded above by the global maximum (\ref{eq:constrMDP}), it must converge to a limit, and depending on the initialization $\locStateEncoderVec_{0}$, this limit may or may not be the global maximum.  When the sequence of control maps and quantizations also converges it must at least be to a Nash equilibrium as summarized in the following theorem.

\begin{theorem}\label{thm:mybigthm}
The iterative method for solving the constrained MDP (\ref{eq:constrMDP}) that is described by  (\ref{eq:ord1}) or (\ref{eq:greedy1}) and (\ref{eq:ord2}),(\ref{eq:ord3}),and (\ref{eq:quantizationUpdate}) yields a monotone increasing sequence of expected rewards which converges.  Additionally, when the sequence of control maps and quantizations converges, $(\controlMap_{*},\locStateEncoderVec_{*}) = \lim_{\iterIdx \rightarrow \infty}  (\controlMap_{\iterIdx},\locStateEncoderVec_{\iterIdx})$, the convergent pair $(\controlMap_{*},\locStateEncoderVec_{*})$ are a Nash equilibrium \cite{MyersonGameTheory}, in the sense that no unilaterial deviation in any of the axes $\locStateEncoderVec$ or $\controlMap'(\possibleMess)$ for each $\possibleMess \in \locStateEncoderVec(\globStateSet)$, can yield an increase in the expected reward.
\end{theorem}

\begin{figure*}
	\centering
	\subfloat[]{\includegraphics[width=0.5\textwidth]{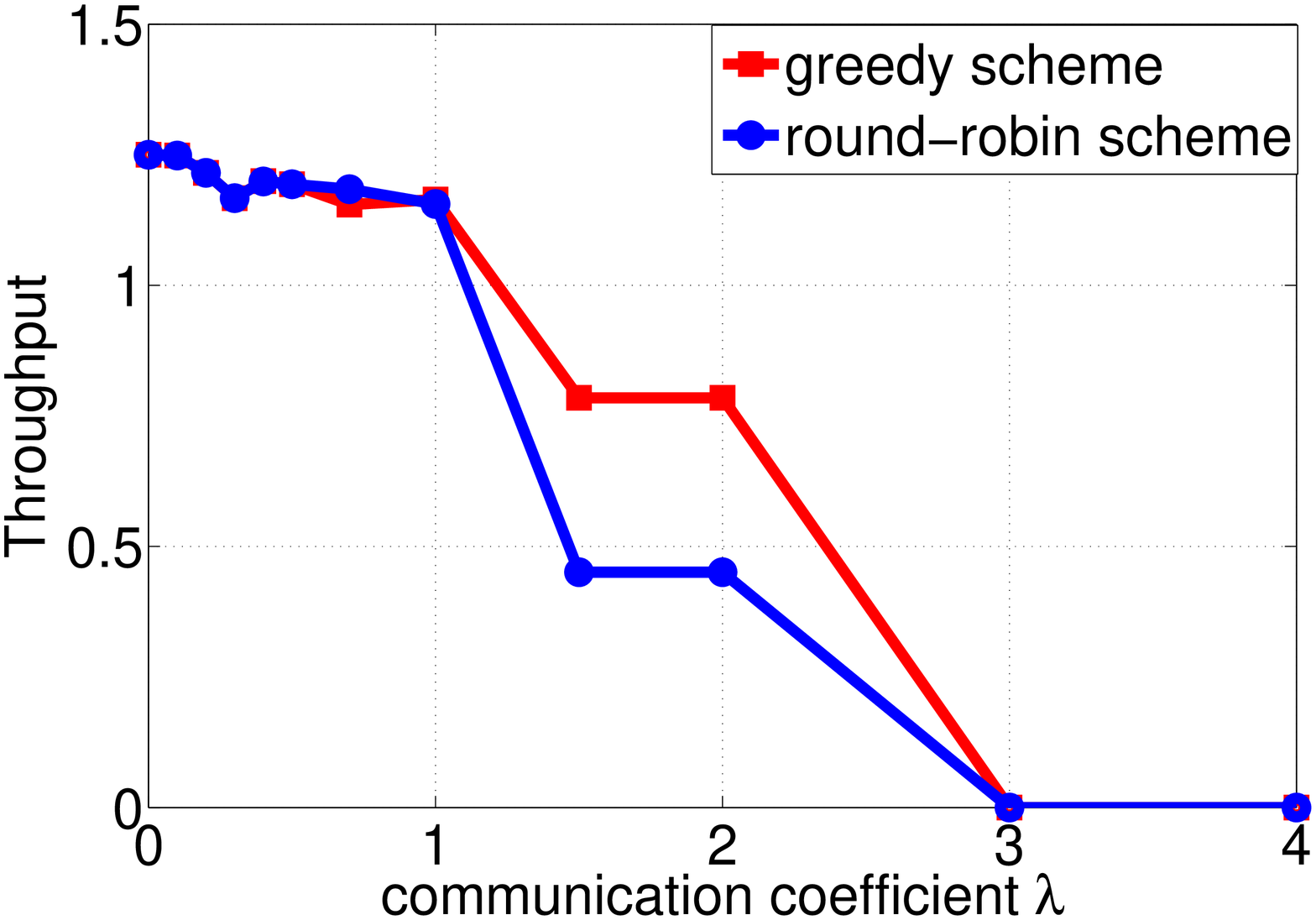}
	\label{fig:throughput-lambda}}
	\subfloat[]{\includegraphics[width=0.5\textwidth]{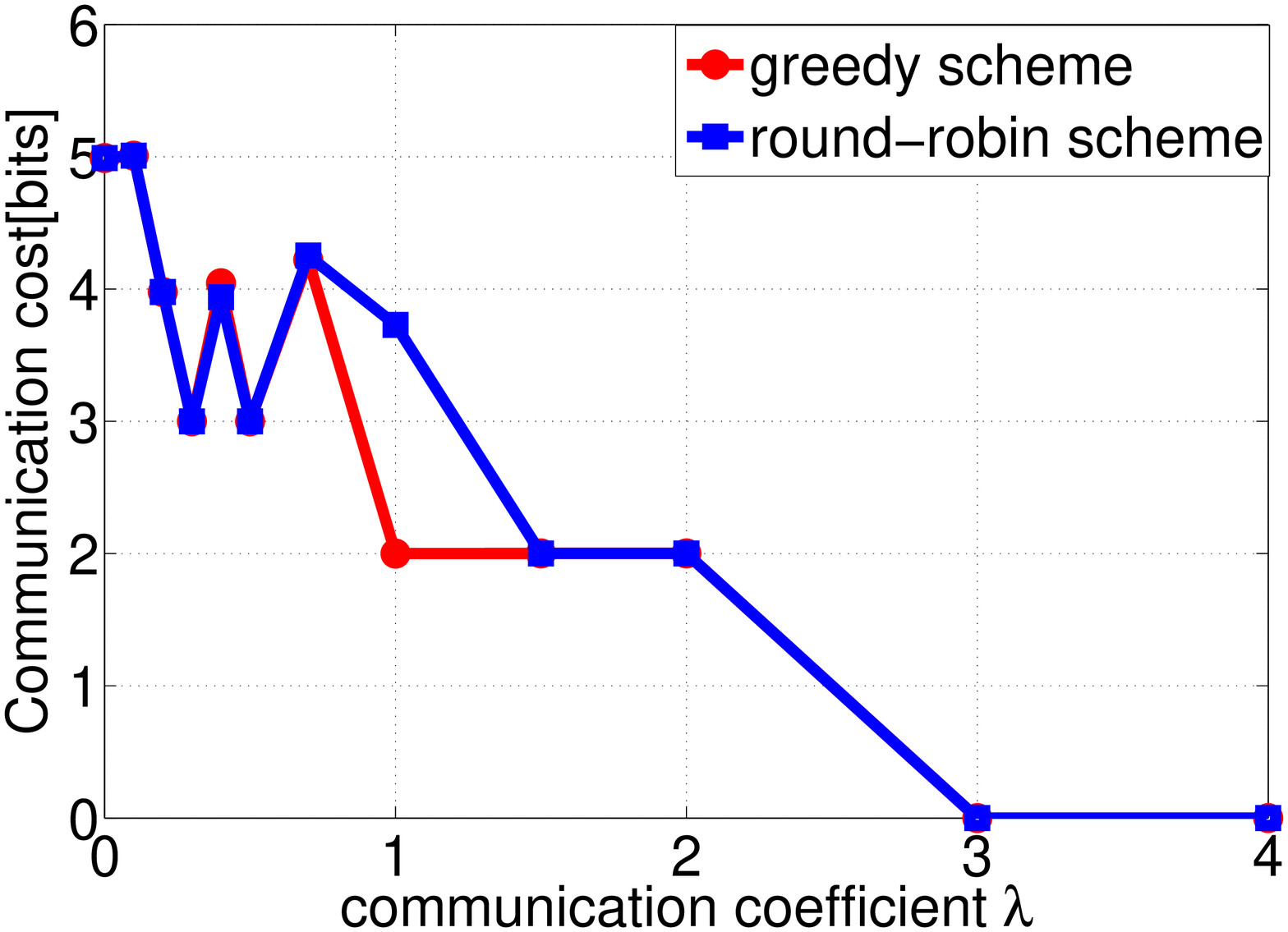}
	\label{fig:cost-lambda}}\\
	\subfloat[]{\includegraphics[width=0.5\textwidth]{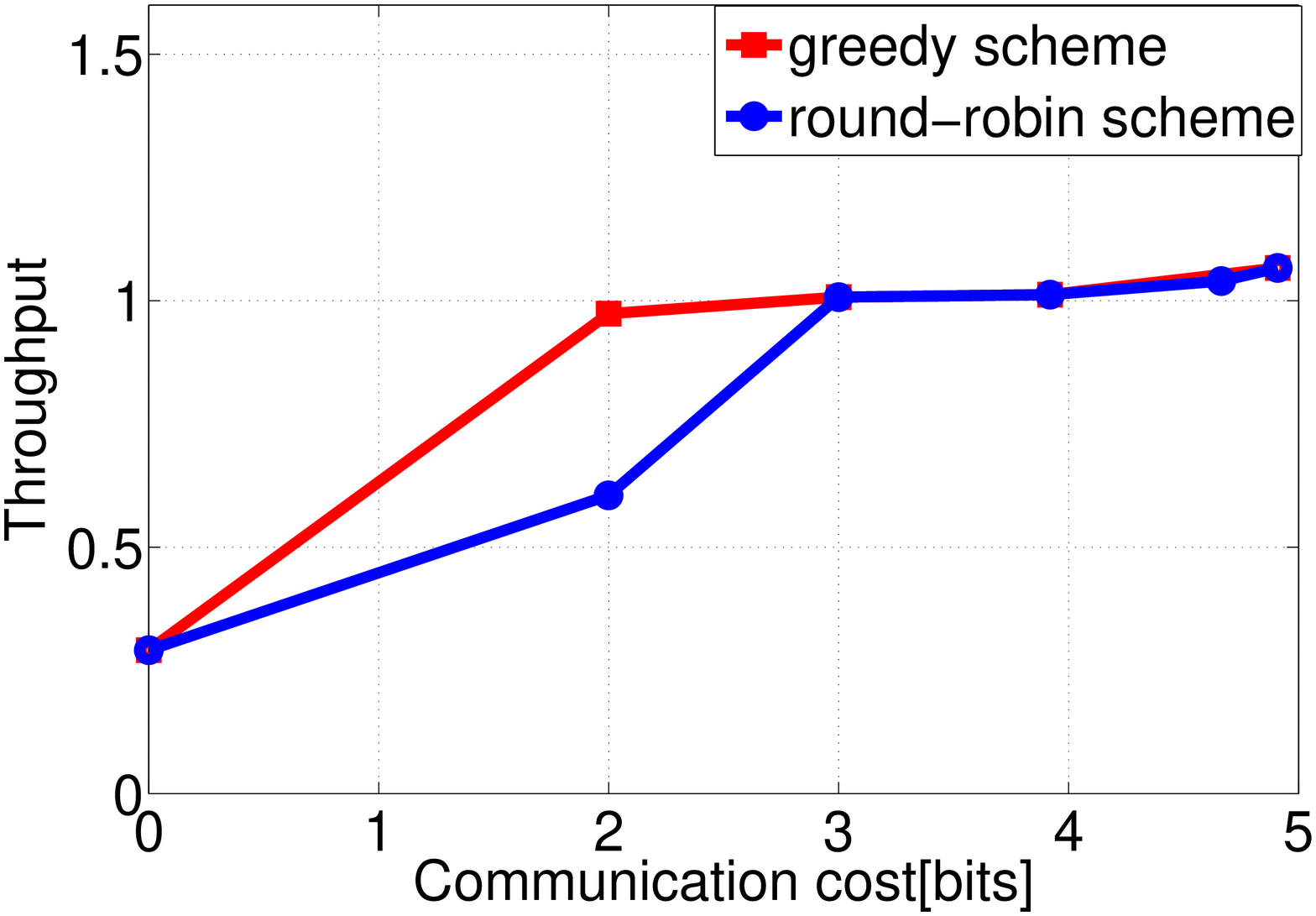}
	\label{fig:throughput-cost}}
	\subfloat[]{\includegraphics[width=0.5\textwidth]{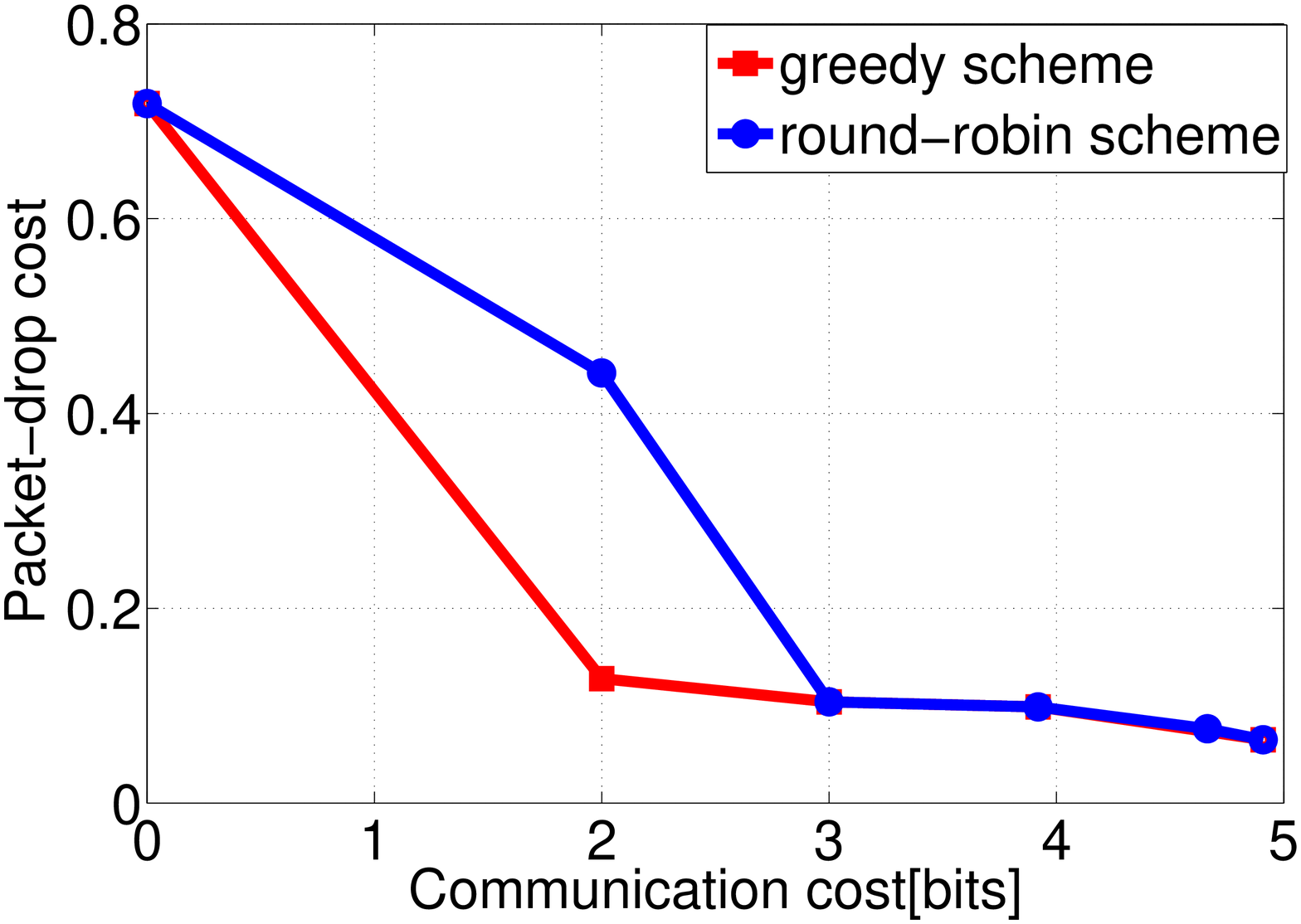}
	\label{fig:packetdrop-cost}}
	
	\caption[]{(a) The expected system throughput, and (b) the control overheads per time-slot with respect to the weighting coefficient $\controlRateLagrMul$ in (\ref{eq:augmented-reward}); (c) The control overheads versus the expected throughput tradeoff, and (d) The control overheads versus the expected packet dropping cost tradeoff computed by both the round-robin and the greedy alternating optimization algorithm with $\controlRateLagrMul \in [0, 4]$ in the augmented reward function in (\ref{eq:augmented-reward})}
	\label{fig:alternating}
\end{figure*}

\begin{example}[Overhead Performance Tradeoff for Wireless Resource Allocation, Larger Model]
We apply the aforementioned alternating optimization algorithm to solve the following example.  Consider again the wireless resource allocation setup of examples \ref{exmp:resourceallocation1} and \ref{exmp:tinyexact}, but in which $\numNodes$($\numNodes = 2$) mobile users observe their local channel quality $\locState_{\nodeIdx,\timeIdx}$ at time-slot $\timeIdx$, distributed on the support $\{0,1,2,3,4\}$ with probabilities $\{\nicefrac{1}{8},\nicefrac{2}{8},\nicefrac{3}{8},\nicefrac{1}{8},\nicefrac{1}{8}\}$ $\forall \nodeIdx \in \{1,2\}, \timeIdx \in \{1,\ldots\}$.  Additionally, the amount of additional traffic arrivals destined for each user, $X_{\nodeIdx,\timeIdx}$ be distributed on the support $\mathcal{X} = \{0,1,2\}$ with probabilities $\{\nicefrac{1}{2},\nicefrac{1}{3},\nicefrac{1}{6}\}$.  The basestation observes the buffer state $\locState_3$ which has a buffer limit $BU_{max} = 4$, and let the packet dropping process be the same as the one described in Example \ref{exmp:one-shot}. 

We refer to the algorithm consisting of (\ref{eq:ord1}),(\ref{eq:ord2}), (\ref{eq:ord3}), and (\ref{eq:quantizationUpdate}) as the round-robin alternating optimization algorithm, and the algorithm consisting of (\ref{eq:greedy1}), (\ref{eq:ord2}), (\ref{eq:ord3}), and (\ref{eq:quantizationUpdate}) as the greedy alternating optimization algorithm.   For the quantizers and information sharing strategies to learn the action, we assume that a one-shot, non-interactive scheme must be used.  Furthermore, both algorithms are initialized with the trivial quantizers which simply relay the full local state.

Note that, as observed by Thm. \ref{thm:mybigthm}, while these alternating optimization methods will always yield a sequence of rewards which is monotone non-decreasing and converges, when the associated control map and quantizer converge, they will in general only be to a Nash equilibrium, and possibly not a global optimum.  This local convergence was indeed observed in the experiments, as some multipliers $\controlRateLagrMul$s lead quantization and control mappings with negative expected discounted rewards, while when not sending any control information, the expected reward will be lower bounded by $0$.  Hence, while the quantization and control schemes presented in the remainder of this example are guaranteed to be Nash equilibria, there is a chance that they are not globally optimal.  Nonetheless, they are highly optimized, and, thus, the tradeoffs they yield, by varying $\controlRateLagrMul$ between the expected throughput and the control overhead are quite interesting.

By applying both the round-robin and the greedy algorithm, we find local optimal quantizations and control mappings for different choice of $\controlRateLagrMul$. 
The expected throughput, and the communication cost are computed as shown in Fig. \ref{fig:throughput-lambda} and \ref{fig:cost-lambda}. 
Based on the local optimal solutions we found, we also plot the throughput versus control overheads tradeoff and the packet dropping cost versus control overheads tradeoff in Fig. \ref{fig:throughput-cost} and \ref{fig:packetdrop-cost}.

We observe from Fig. \ref{fig:throughput-lambda} that the expected system throughput is maximized when $\controlRateLagrMul = 0$. In such a case, the communication cost is not involved in the transition reward function, the optimal control mapping $\controlMap$  becomes to pick the user that maximize the instant throughput, and the optimal encoders $\locStateEncoderVec$ are designed to minimize the control overheads while guarantee that the optimal control mapping can be learned by any node with only observing all of the control messages. 
 We also observe from Fig. \ref{fig:cost-lambda} that the control overheads become $0$ when $\controlRateLagrMul \ge 3$, this is because, as $\controlRateLagrMul$ grows larger, the system realizes that the cost it pays to encode the local states weights more than the reward it could earn by sending data traffics to the destined user, hence the optimal decision is to not encode any local state and blindly schedule a user to occupy the resource block.  
Finally, we observe from Fig. \ref{fig:throughput-lambda} that the expected system throughput per time-slot goes to $0$ when the system decision is to blindly pick a user. This is because, although the basestation may pick users and send data traffics in the first few time-slots, however, in a long run, the global state must be absorbed to one of the recurrent classes in which, the instant throughput will always remain $0$, as the buffer fills with traffic destined for the other users and stays full. In fact, given a blind control mapping with no local observation as
 \begin{equation}
 \controlMap(\globState) = \actiona,\ \text{for all}\ \globState \in \mathcal{S},
 \end{equation}
 those global states with the same buffer local state $\locState_{\numNodes+1} = (b_1,\ldots,b_{\numNodes})$ satisfying 
 \begin{equation}
 b_a = 0
 \end{equation}
 and
 \begin{equation}
 \sum_{\nodeIdx \in \nodeSet  \setminus \actiona} b_{\nodeIdx} = BU_{max} 
 \end{equation}
will form a recurrent class. The result that the expected system throughput becomes $0$ when the system decision is to blindly pick a user indicates that a system with randomly picking users at all time-slots will perform better than the deterministic control mapping system, which matches the conclusion in \cite{SJJ94}\cite{MurphyPOMDP}.  It is important to note, however, that this would be precluded by the present model, which would require which user would be transmitted to be known deterministically to a participant just arriving in the network who in this case would not have observed anything since there is no control information being sent.  Additional control rate savings and increased rewards enabled by randomization, which will require the assumption of synchronized common randomness at all participants in the scheme, are an important direction for future work.
%
\end{example}
\section{Conclusion}
\label{sec:conclusion}
This paper analyzed a Markov decision process in which the state was composed of a series of local states, each observed at a different location in a network.  Using recent results from multiterminal information theory regarding distributed and interactive function computation, the minimum amount of control information that would be necessary to exchange in order for the system to simulate a centralized controller having access to the global state was determined.  Next, the information theoretic cost of communication was incorporated into the reward function in the MDP, and the problem of simultaneously designing the controller and the messaging scheme to maximize the associated combined reward was formulated, creating a tradeoff between communication and performance.  To provide candidate solutions for the associated optimization problem, an alternating optimization method was presented that produces a sequence of rewards that always converges, and when the associated messaging scheme and controller map converges, it converges to a Nash equilibrium for the problem.  A series of running examples from downlink wireless resource allocation illustrated the ideas throughout.  Important directions for future investigation involve allowing time varying messaging and control schemes, the use of historical observations of messages in a POMDP like framework, and the use of rate distortion theory to aid with the derivation of tradeoffs between communication and control reward in the present decentralized MDP context.

\bibliographystyle{IEEEtran}
\bibliography{sources}

\begin{IEEEbiography}[{\includegraphics[width=1in,height=1.25in,clip,keepaspectratio]{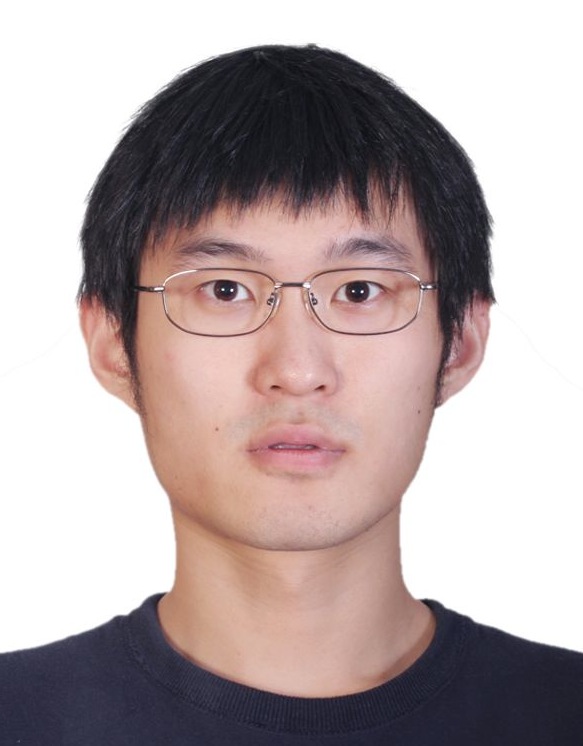}}]{Jie Ren} (S'14)
received the B.S. degree in Electrical Engineering from Tsinghua University, China, in 2009, and the M.S. degree in Telecommunication from Widener University, Chester, PA in 2011. Since September 2011 he has been pursuing a Ph.D. at Drexel University, Philadelphia, PA within the Adaptive Signal Processing and Information Theory Research Group.  After interning with the Wireless Big Data Research Team, Huawei(US) in 2015, he joined Futurewei (Bridgewater, NJ) in 2016 as a software engineer in wireless access.
\end{IEEEbiography}

\begin{IEEEbiography}[{\includegraphics[width=1in,height=1.25in,clip,keepaspectratio]{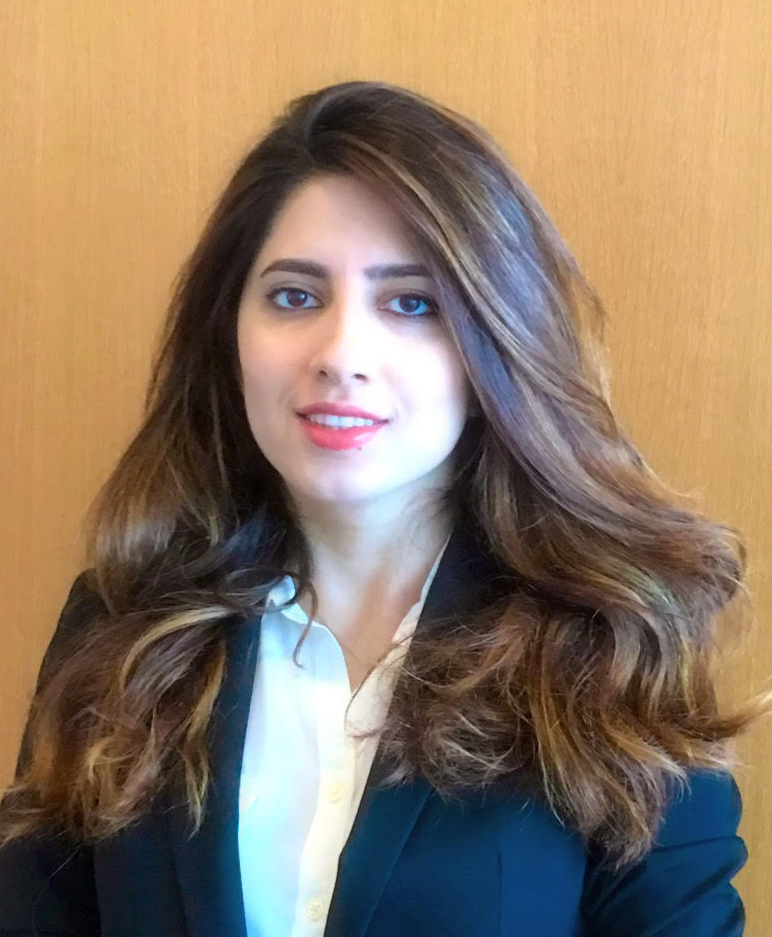}}]{Solmaz Torabi} (S'14)
received the B.Sc. degree in Electrical Engineering from Sharif University of Technology, Iran, in 2011. She has been working toward her Ph.D. degree since 2013 within the Adaptive Signal Processing and Information Theory Research Group in the department of Electrical and Computer Engineering at Drexel University, Philadelphia, PA. Her research interests include the areas of information theory and interactive communication.
\end{IEEEbiography}

\begin{IEEEbiography}[{\includegraphics[width=1in,height=1.25in,clip,keepaspectratio]{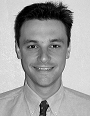}}]{John MacLaren Walsh} (S'01-M'07-SM'15) received the B.S. (\emph{magna cum laude}), M.S. and Ph.D. degrees in electrical and computer engineering from Cornell University, Ithaca, NY in 2002, 2004, and 2006, respectively.  In September 2006, he joined the Department of Electrical and Computer Engineering at Drexel University, Philadelphia PA, where he is currently a associate professor.  At Drexel, he directs the Adaptive Signal Processing and Information Theory Research Group. 
\end{IEEEbiography}

\end{document}